\documentclass{ws-ijmpe}
\usepackage[super,compress]{cite}
\usepackage{upgreek}
\usepackage{float}
\usepackage{subfigure}
\usepackage{graphicx}
\begin{document}

\markboth{W. Poonsawat, C. Kobdaj, M. Sitta, Y. Yan}{Material Budget Calculation of the new Inner Tracking System, ALICE}

\catchline{}{}{}{}{}

\title{Material Budget Calculation of the new Inner Tracking System, ALICE}

\author{WANCHALOEM POONSAWAT}

\address{School of Physics, Institute of Science, Suranaree University of Technology\\
Nakhon Ratchasima 30000, Thailand\\
wanchaloem.poonsawat@cern.ch}

\author{CHINORAT KOBDAJ\footnote{Corresponding author}}

\address{School of Physics, Institute of Science, Suranaree University of Technology\\
Nakhon Ratchasima 30000, Thailand\\
kobdaj@g.sut.ac.th}

\author{MARIO SITTA}

\address{Universit$\acute{a}$ del Piemonte Orientale, Alessandria and INFN\\
Torino 10125, Italy\\
sitta@to.infn.it}

\author{YUPENG YAN}

\address{School of Physics, Institute of Science, Suranaree University of Technology\\
Nakhon Ratchasima 30000, Thailand\\
yupeng@g.sut.ac.th}

\maketitle

\begin{history}
\received{Day Month Year}
\revised{Day Month Year}
\end{history}

\begin{abstract}
The ALICE Collaboration aims at studying the physics of strongly interacting matter by building up a dedicated
heavy-ion detector. The Inner Tracking System (ITS) is located in the heart of the ALICE Detector surrounding the interaction point. Now, ALICE has a plan to upgrade the inner tracking system for rare probes at low transverse momentum. The new ITS composes of seven layers of silicon pixel sensor on the supporting structure. One goal of the new design is to reduce the material budget ($X/X_0$) per layer to 0.3$\%$ for inner layers and 0.8$\%$ for middle and outer layers. In this work, we perform the simulations based on detailed geometry descriptions of different supporting structures for inner and outer barrel using ALIROOT. Our results show that it is possible to reduce the material budget of the inner and outer barrel to the value that we have expected. The manufacturing of such prototypes are also possible.
\end{abstract}


\ccode{PACS numbers: 29.40.Gx, 29.90.+r}


\section{Introduction}

A Large Ion Collider Experiment (ALICE) is designed to studies the property of Quark-Gluon Plasma (QGP), the deconfined state of strongly interacting matter. There are eighteen systems within ALICE detector. In this work, we are interested in a central part of ALICE detector called the Inner Tracking System. The present ITS is made of six layers exploiting three different silicon technologies. The main duty of ITS is to detect the primary (where the collision occurs) and secondary vertices (where some of the unstable heavy particles decay after a flight distance of some hundreds of micrometers). The innermost layers are needed to be high resolution devices to record both x and y coordinates for each passing particle.\cite{Evans}

The precision of the present ITS for charm mesons is insufficient at low transverse momentum ($<$ 1 GeV/c) and even worse for charm baryons. In case of charm baryon, the lowest-mass charm baryon is the $\Lambda_c$ with a rest mass of 2286.46 $\pm$ 0.14 MeV/c$^2$.\cite{Nakamura:1299148} The important channel of its measurement is the decay of $\Lambda_c \to pK^-\pi^+$ with a branching ratio equal to 5.0 $\pm$ 1.3 $\%$. The mean proper decay length ($c\tau$) of $\Lambda_c$ is only 60 $\upmu$m with a very short lifetime that turns out to be shorter than the impact parameter resolution of the present ITS. Therefore, the charm baryons are partially accessible by the ALICE detector in central Pb-Pb collisions. 

The current measurements in Pb-Pb collisions are characterized by a very small signal-over-background ratio, which calls for large statistics. Thus, to reach the goal of heavy quark measurement, the ALICE ITS must be upgraded.\cite{musa}

\section{Current and upgrade Inner Tracking System}

The current ITS is composed of six concentric barrels with three different technologies of detector surrounding the beam-pipe. The Silicon Pixel Detectors (SPD) are used in the first two innermost layers. The next two layers are made of Silicon Drift Detectors (SDD) and the two outermost layers are equipped with the Silicon Strip Detector (SSD).\cite{musa} 

Due to the upgraded beam-pipe radius reduction from 2.94 cm to 1.98 cm, this allows us to place the innermost layer 9.6 mm closer to the interaction point. An extra layer is also introduced to increase tracking efficiency of the new ITS. This new design is expected to improve the impact parameter resolution by a factor of three~\cite{musa}. The use of Monolithic Active Pixel Sensors (MAPS) for the seven layers of ITS upgrade can reduce the material budget by a factor of 7 (50 $\upmu$m instead of 350 $\upmu$m) per layer.\cite{LEMMON} For data readout, the analog front-end timing and data transfer architecture will be developed to increase frequency greater than 50 kHz for Pb-Pb interaction and about 7 MHz for p-p interaction.\cite{Rossegger201340} The new cooling system also is improved for a better heat transfer.

\begin{figure}[th]
\centerline{\includegraphics[width=4.in]{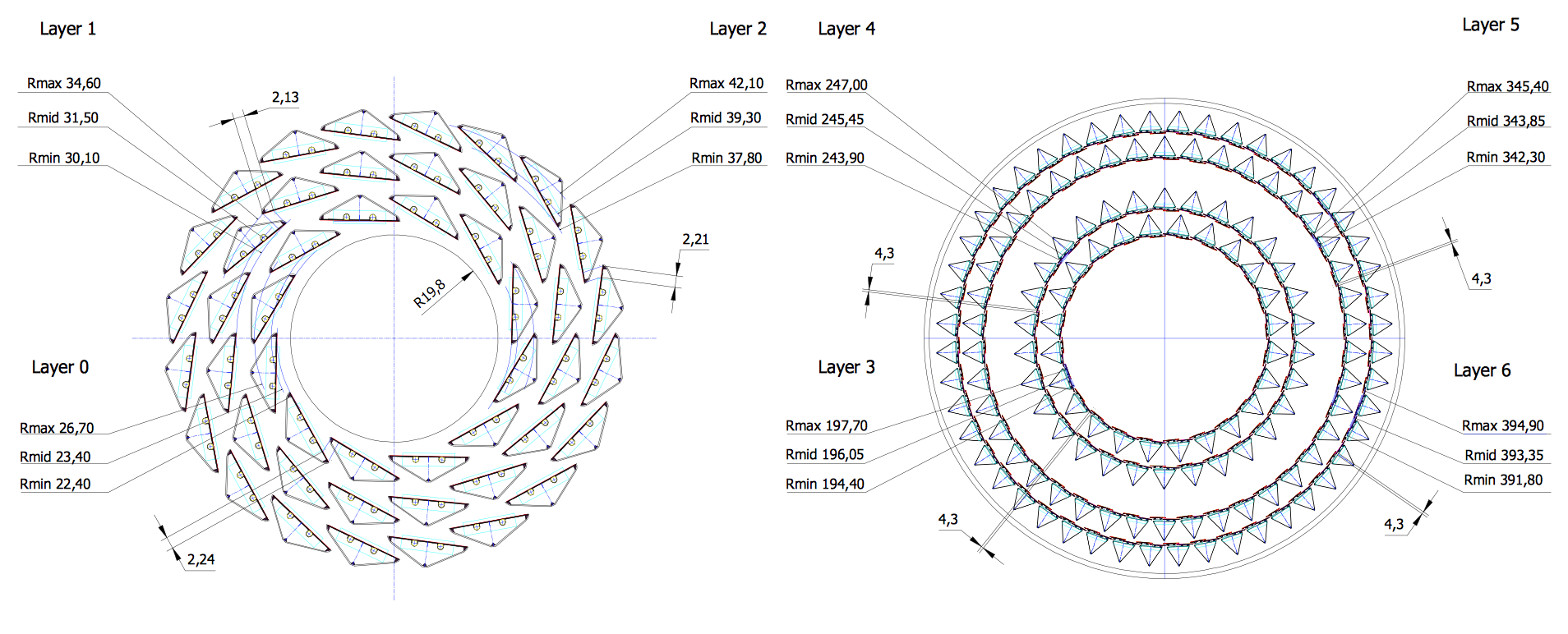}}
\caption{\label{layer} Schematic cross section layout of the inner barrel (left) and outer barrel (right).\cite{abelev}}
\end{figure}

The design of the new ITS composes of the inner barrels (layers 0 to 2), the middle barrels (layers 3 to 4) and the outer barrels (layers 5 to 6) (see in Fig.~\ref{layer}).

The layout of the new design are shown in Table~\ref{tab0} with the expected material budget.

\begin{table}[pt]
\tbl{\label{tab0} The layout of the upgrade scenario for the ITS inner barrel. The numbers in brackets refer to the case of current detector.}
{\begin{tabular}{@{}ccc@{}} \toprule
Layer & Radius (cm)& Material budget ($\%X_0$)\\ \colrule
inner barrel & & \\
 1 pixel (pixel) & 2.2 (3.9) & 0.3 (1.14)\\
 2 pixel (pixel) & 2.8 (7.6) & 0.3 (1.14)\\
 3 pixel (none) & 3.6 (-) & 0.3 (-)\\
mid barrel & & \\
 4 pixel (driff) & 20.0 (15.0) & 0.8 (1.13)\\
 5 pixel (driff) & 22.0 (23.9) & 0.8 (1.26)\\
outer barrel &  & \\
 6 pixel (strip) & 41.0 (38.0) & 0.8 (0.83)\\
 7 pixel (strip) & 43.0 (43.0) & 0.8 (0.83)\\ \botrule
\end{tabular}}
\end{table}

\subsection{The new stave structure of upgrade ITS}

The $\textit{stave}$ is a composite element of the upgraded ITS layers devised to be a support structure to be a support structure. It consists of the following  components:

\begin{itemlist}
\item Space Frame: a lightweight wound truss structure made of carbon fibre for mechanical supporting.    
\item Cold Plate: a carbon ply with  embedded cooling units.
\item Hybrid Integrated Circuit: an electronic circuit with pixel sensor chip.
\end{itemlist}

In this work, both the inner and outer barrels have been studied and their material budgets have been calculated. The first three inner layers are built with the identical staves. The space frame of each stave is made of a light filament wound carbon. It is obtained by winding a M60J 3K (588GPa) carbon rowing respect to the stave axis with an angle of 45$^\circ$. The winding angle and the number of helices have been optimized to achieve the best compromise between material budget and stiffness.

For the cooling system, cold plate is used to remove the heat dissipated from the pixel chips. The cold plate is made of a high thermal conductive carbon fibre laminated on top of which the silicon chips are glued. The heat is conducted into the cooling pipes or microchannel embedded in the cold plate and is removed by the coolant. In order to maximise the cooling efficiency, the cold plates have been considered in four different models for the various geometry design and the thermal constraints.

The mechanical support structure is designed under the technical constraints such as the requirement of detector layout, specific sensor chip size and the cooling technology. The stave structures have been separated into several models. 

\begin{itemlist}
\item Model 0: The wound truss structure with cooling pipes at the vertices
\end{itemlist}

\begin{figure}[th]
\centering
\subfigure[Model 0]{%
  \includegraphics[width=2.in, height=1.in]{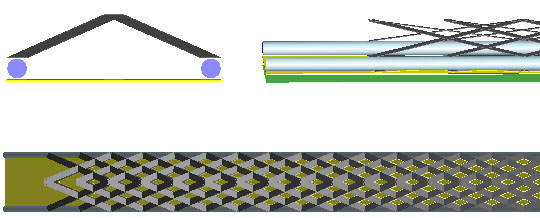}
  \label{fig:subfigure1}}\hspace{-2em}%
\quad\quad
\subfigure[Model 1]{%
  \includegraphics[width=2.in, height=1.in]{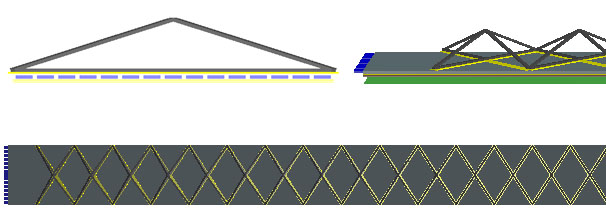}
  \label{fig:subfigure2}}
  \subfigure[Model 2.1]{%
  \includegraphics[width=2.in, height=1.in]{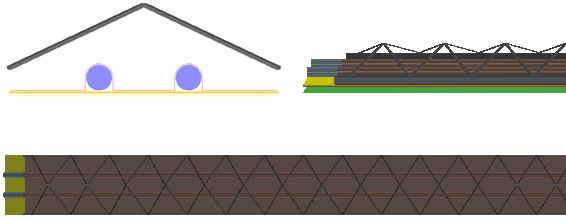}
  \label{fig:subfigure3}}\hspace{-2em}%
\quad\quad
\subfigure[Model 2.2]{%
  \includegraphics[width=2.in, height=1.in]{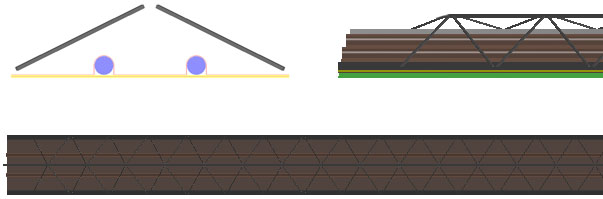}
  \label{fig:subfigure4}}
  \subfigure[Model 3]{%
  \includegraphics[width=2.in, height=1.in]{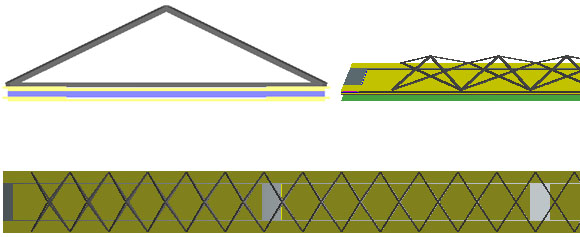}
  \label{fig:subfigure5}}\hspace{-2em}%
\quad\quad
\subfigure[Model 4]{%
  \includegraphics[width=2.in, height=1.in]{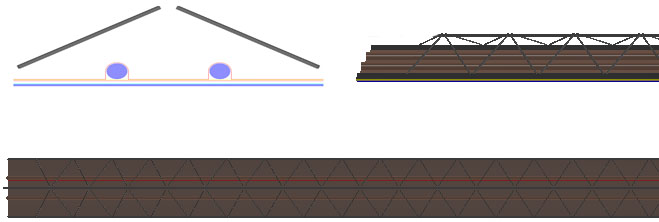}
  \label{fig:subfigure6}}
\caption{\label{fig:figure} \small Assembly of the inner barrel stave prototype composed of space frame, cooling structure and sensor chip. For material analysis the different structures are being considered a) The wound truss structure with cooling pipes at the vertices, b) The wound truss structure with polyimide microchannel cooling, c) The wound truss structure with uniform carbon plate and 0.15 mm outer radius polyimide tubes in the middle, d) The wound truss structure with uniform carbon plate and 0.10 mm outer radius polyimide tubes in the middle, e) The wound truss structure with silicon microchannel and f) The wound truss structure with uniform carbon plate and polyimide tubes in the middle (revisited).}
\end{figure}

The winding carbon filaments with 5 mm diameter are used to construct the support structure. K13D2U is used for wound truss structure instead of M55J 6K carbon fiber for a better thermal conductive of two embedded pipes as shown in Fig.~ \ref{fig:subfigure1}. The use of K13D2U carbon type can reduce the bending radius against M55J 6K. The carbon fiber are wound with an angle of 23 degrees along the stave axis to reduce the breaking of production process. This prototype can reduce the weight of structure while preserving a good stiffness. The pyramidal structure made by the carbon fiber filaments will be strong enough to support the silicon  sensors and the layer structure.  The overall structure is suitable with low power consumption.

\begin{itemlist}
\item Model 1: The wound truss structure with polyimide microchannel cooling
\end{itemlist}

For this model, the same technique described for wound truss structure has been used but the cooling pipes have been replaced by a cold plate for a better distributed heat transferring over the pixel chip. A monophase or biphase refrigerant fluid is used in the cold plate flowing into 0.16 mm$^2$ of polyimide microchannel cooling section (see in Fig.~\ref{fig:subfigure2}). 

\begin{itemlist}
\item Model 2: The wound truss structure with uniform carbon plate and polyimide tubes in the middle
\end{itemlist}

This model, obtained with the same technology described above, uses a uniform carbon plate for transferring heat to two embedded tubes. These polyimide tubes are placed in the middle at the base of the stave structure. However, the addition of a carbon plate may cause the increase of material.

In this model, two different outer radii, 0.15 mm and 0.10 mm, of cooling pipes are separately simulated for the material properties. 

\begin{itemlist}
\item Model 3: The wound truss structure with silicon microchannel
\end{itemlist}

This model is the wound truss structure with cold plate, made of a silicon substrate (see Fig.~\ref{fig:subfigure5}). Based on etching technology, the microchannel can be created on the silicon plate where the cooling fluid flows inside. The microchannels are fabricated in a clean room of the Center of Micronanotechnology, Ecole Polytechnique Federale de Lausanne and Thai Microelectronics Center. Due to some complication in its fabrication process, the prototype is just being developed with substantial cost.

\begin{itemlist}
\item Model 4: the wound truss structure with uniform carbon plate and polyimide tubes in the middle (revisited)
\end{itemlist}

The concept of this model is similar to Model 2 with 0.10 mm of outer radius but the structures were completely rewritten to make them more systematic and similar to those of in the outer barrel. Some inconsistencies such as the element sequence and empty space are corrected (see in Fig.~\ref{fig:subfigure6}).

By coding a detailed geometry description of all models in ALIROOT, the complete calculation of material budget of all stave models can be performed.

All models have been considered and proceeded to calculate their material budget distribution. Remarks and comments for each models have been passed to the prototype production in the ALICE ITS upgrade.

\section{Toward the material budget ($X/X_0$) reduction of inner barrel}

The energy lost of a particle passing through matter is related to properties of materials. The radiation length ($X_0$) is the mean path length required to reduce the energy of relativistic charged particles by the factor $1/e$ related to its energy loss~\cite{Agashe:2014kda}. The radiation length of material can be approximated by the expression of the atomic number and atomic weight of the nucleus. In case of compound material or mixture, the radiation length can be estimated with the combination of all compound radiation length multiplied by the mass fraction~\cite{Mukund}. The properties of materials used in the ALICE ITS upgrade scenarios are given in Table \ref{tab1}.

\begin{table}[pt]
\tbl{\label{tab1} List of the stave components and their thickness used in ALIROOT simulation. The numbers in brackets refer to the conceptual parameters in CDR.}
{\scriptsize\begin{tabular}{@{}ccccccc@{}} \toprule
Material & Model 0 & Model 1 & Model 2.1 & Model 2.2 & Model 3 & Model 4 \\
& \multicolumn{6}{c}{Thickness [$\upmu$m]}\\ \colrule
\textbf{Filament}\hphantom{00000000000} & & & & & & \\
  \hphantom{00}Top CFRP M60J 3K & 70 (70) & 200 (120) & 200 (100) & 200 (100) & 200 (120) & 200 (100)\\
  \hphantom{00000}Bottom CFRP M60J 3K & 70 (-) & 200 (240) & - & - & 200 (240) & -\\
\hline
\textbf{Cooling}\hphantom{000000000000} & & & & & & \\
Pipe Kapton\hphantom{00000} & 70 (70) & - & 130 (70) & 130 (70) & - & 130 (70)\\
water\hphantom{00000000000} & 1450 (1450) & - & 1450 (1450) & 940 (940) & - & 940 (940)\\
Carbon plate\hphantom{00000} & - & - & 140 (140) & 140 (140) & - & 140 (140)\\
\hline
\textbf{Sensor}\hphantom{000000000000} & & & & & & \\
Glue\hphantom{000000000000} & 125 (200) & 250 (200) & 100 (200) & 100 (200) & 250 (200) & 100 (200)\\
Silicon chip\hphantom{000000} & \multicolumn{6}{c}{50 (50)}\\
Flex cable\hphantom{0000000} & \multicolumn{6}{c}{100 (-)}\\
\hphantom{00000}Polyimide Microchannel & - & 100 (100) &  -  & - & - & -\\
\hphantom{00}Silicon Microchannel & - & - & - & - & 40 (40) & -\\
water\hphantom{00000000000} & - & 200 (200) & - & - & 160 (160) & -\\ \botrule
\end{tabular}}
\end{table}

The material budget of each stave prototypes can be calculated from the radiation length of their components. It either depends on thickness ($X/X_0$) or percentage of covered surface ($X/X_0(\%)$). The computation of the material budget in ALIROOT is done using fake particles, called ``geantino''s. They can be shot straight to the sensors, and having no charge they suffer neither deviation nor any energy loss. They are tracked as all other particles in each step. The user can restrict the volumes crossed by geantinos by setting the minimum radius $R_{\rm{min}}$ and the maximum radius $R_{\rm{max}}$ , the minimum and maximum $\phi$ values, and the range in $Z$, between $−Z_{\rm{min}}$ and $+Z_{\rm{max}}$ of the travelling region. The material budget can be determined in essentially two ways. In the first one the geantino tracks are generated perpendicular to the $Z$ axis of the staves, hence, the ``actual'' material budget can be determined. In the second way, all geantinos come from the interaction point, so the material budget is the same as seen by real particles coming from the collisions depending on the provided $\eta$ range. 

The stave design accounts for material budget requirement which is limited to 0.3$\%$ for the inner barrel and 0.8$\%$ for the outer barrel~\cite{musa}. The simulations include essentially all material properties used in each model as shown in Table \ref{tab1} and Table~\ref{tsum}.

\begin{table}[pt]
\tbl{\label{tsum} List of the stave components and their contribution to the radiation length. The numbers in brackets refer to the approximation value in CDR.}
{\scriptsize\begin{tabular}{@{}lcccccc@{}} \toprule
Material & Model 0 & Model 1 & Model 2.1 & Model 2.2 & Model 3 & Model 4 \\
& \multicolumn{6}{c}{Thickness [$\upmu$m]}\\ \colrule
\textbf{Filament} & & & & & & \\
Top CFRP M60J 3K & \multicolumn{6}{c}{19 (25)}\\
Bottom CFRP M60J 3K & 19 (25) & 19 (25) & - & - & 19 (25) & -\\
\hline
\textbf{Cooling} & & & & & & \\
Pipe Kapton & 28.4 (28.6) & - & 28.4 (28.6) & 28.4 (28.6) & - & 28.4 (28.6)\\
water & 35.8 (36.1) & - & 35.8 (36.1) & 35.8 (36.1) & - & 35.8 (36.1)\\
Amec Thermasol FGS 003 & - & - & 27 (25) & 27 (25) & - & 27 (25)\\
C Fleece & - & - & 106 (25) & 106 (25) & - & 106 (25)\\
\hline
\textbf{Sensor} & & & & & & \\
Glue & \multicolumn{6}{c}{44.37 (44.37)}\\
Silicon chip & \multicolumn{6}{c}{9.35 (9.36)}\\
Flex cable & \multicolumn{6}{c}{13.3 (13.3)}\\
Polyimide Microchannel & - & 28.4 (28.6) &  -  & - & - & -\\
Silicon Microchannel & - & - & - & - & 9.35 (9.36) & -\\
water & - & 35.8 (36.1) & - & - & 35.8 (36.1) & -\\
K13D2U 2K & - & - & 26 (25) & 26 (25) & - & 26 (25)\\
C Fleece & - & - & 106 (25) & 106 (25) & - & 106 (25) \\ \botrule
\end{tabular}}
\end{table}

\begin{figure}[th]
\centerline{\includegraphics[width=3.8in]{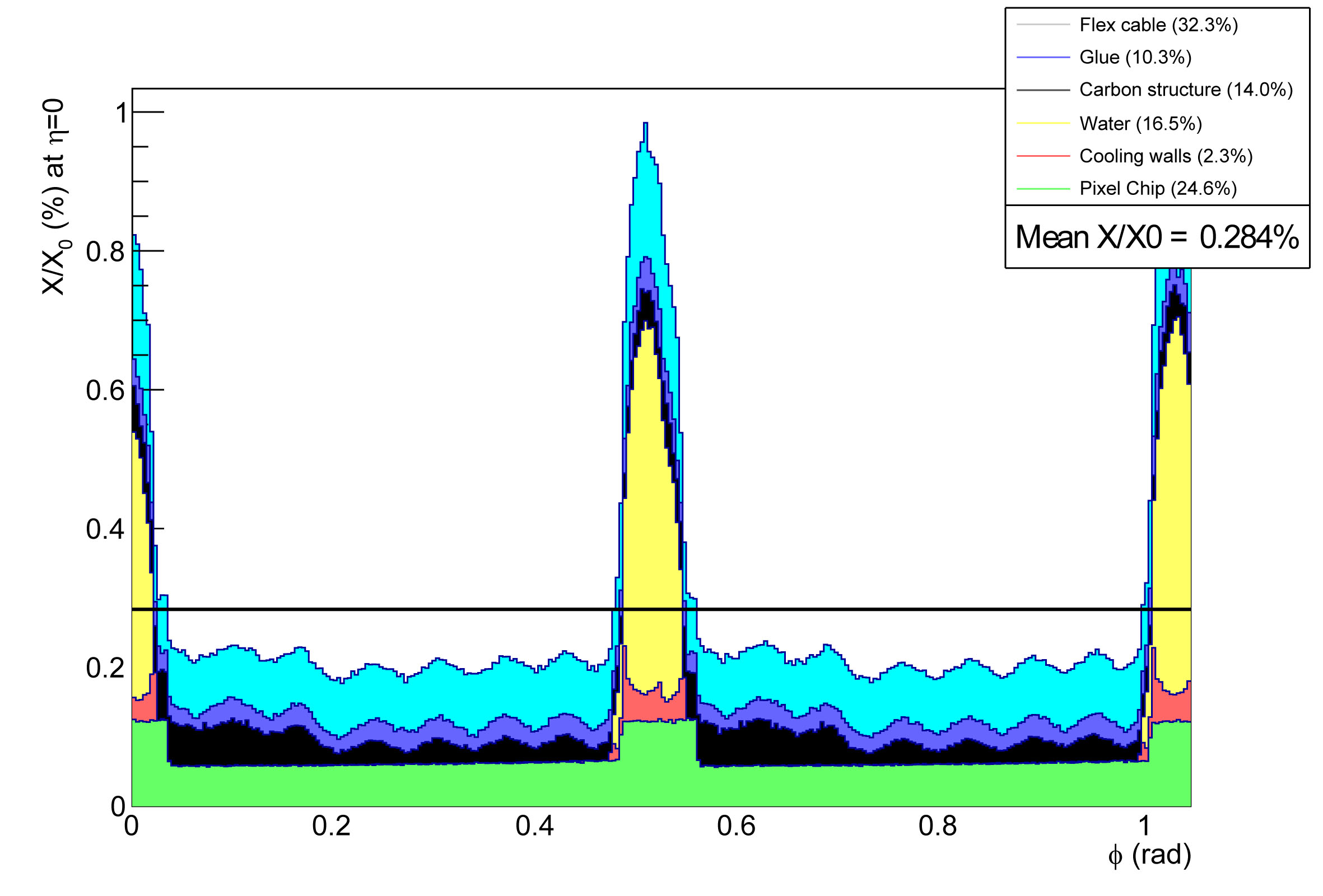}}
\caption{\label{mat0} The material budget distribution of Model 0.}
\end{figure}

\begin{figure}[th]
\centerline{\includegraphics[width=3.8in]{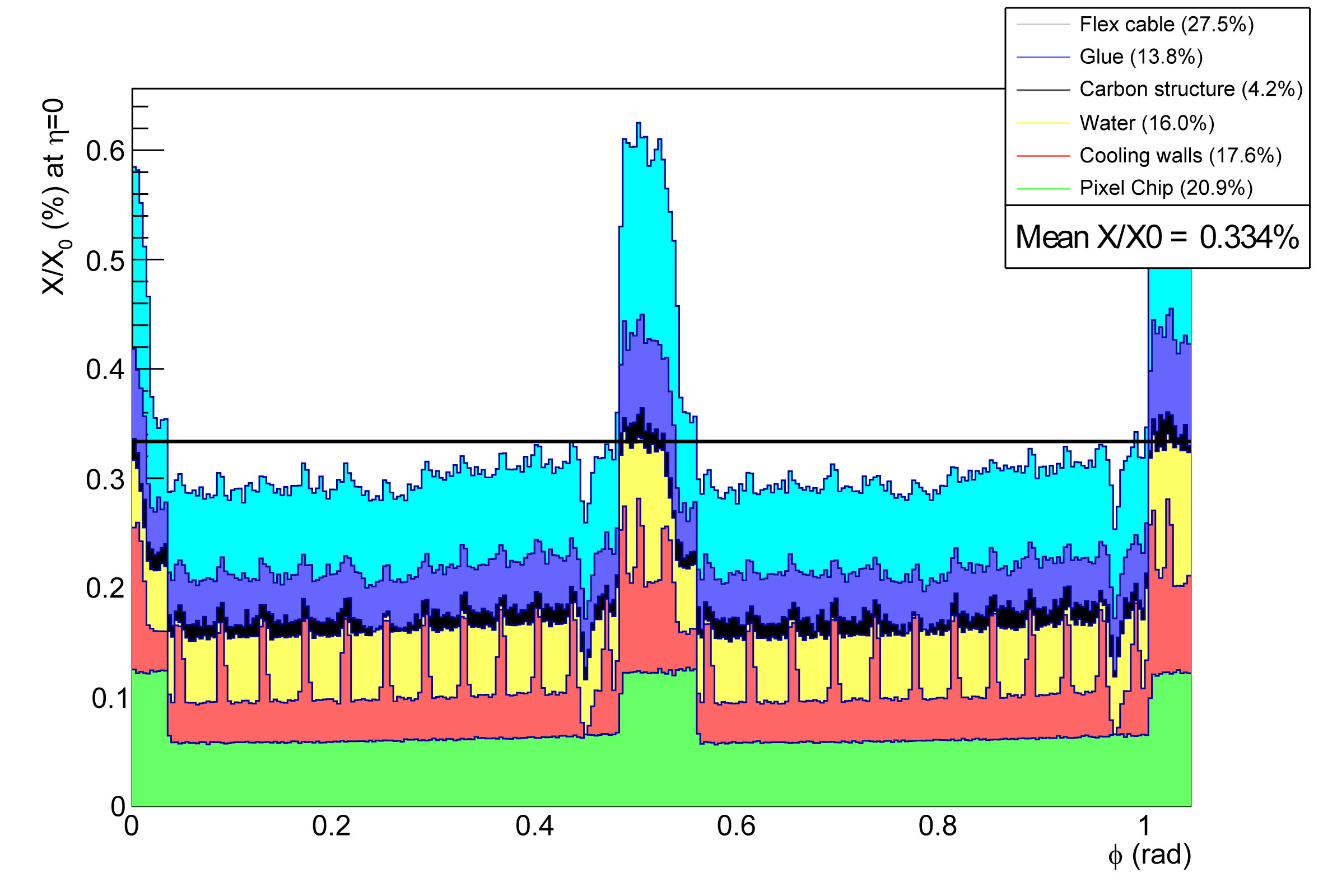}}
\caption{\label{mat1} The material budget distribution of Model 1.}
\end{figure}

\begin{figure}[th]
\centerline{\includegraphics[width=3.8in]{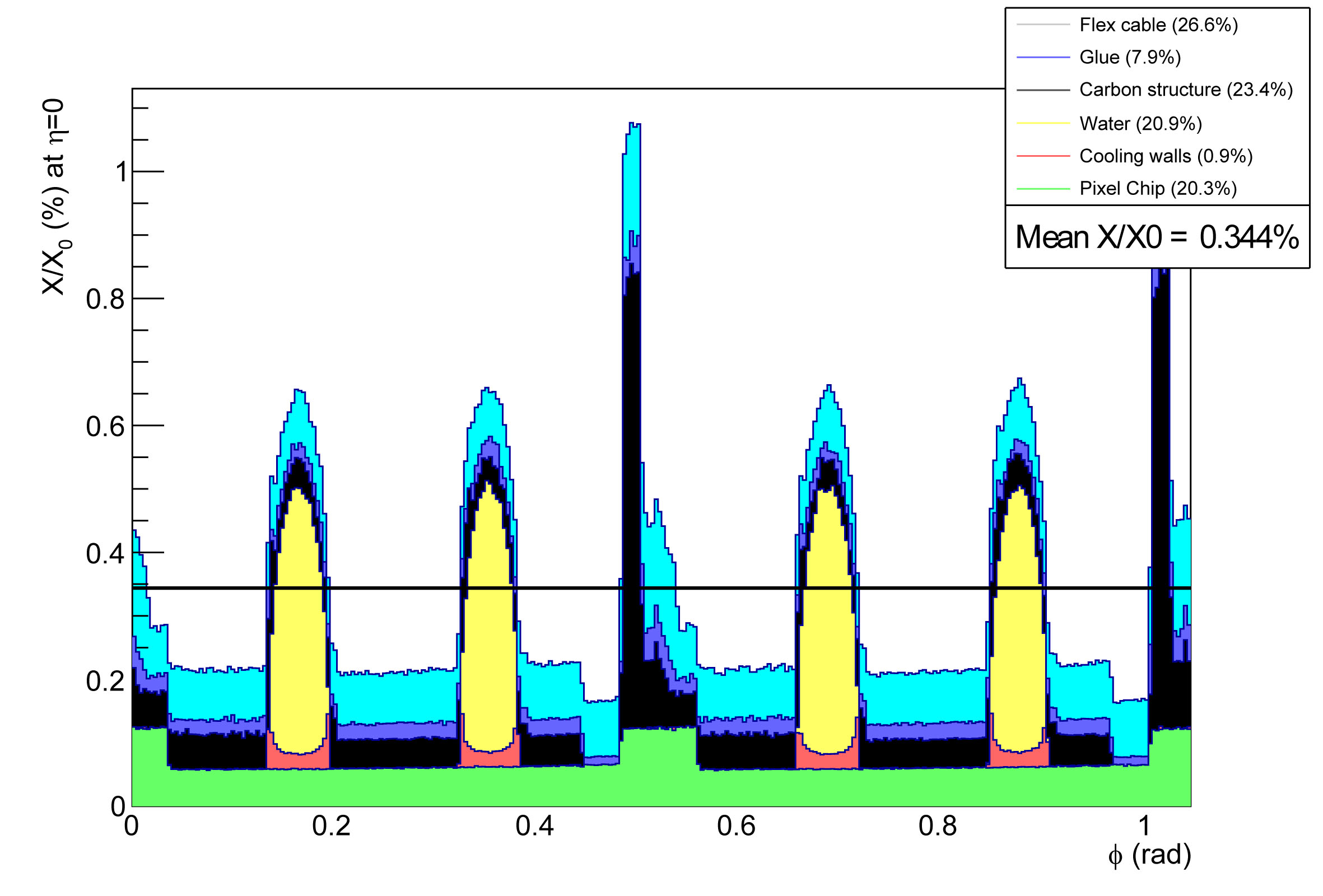}}
\caption{\label{mat21} The material budget distribution of Model 2.1.}
\end{figure}

\begin{figure}[H]
\centering{\includegraphics[width=3.8in]{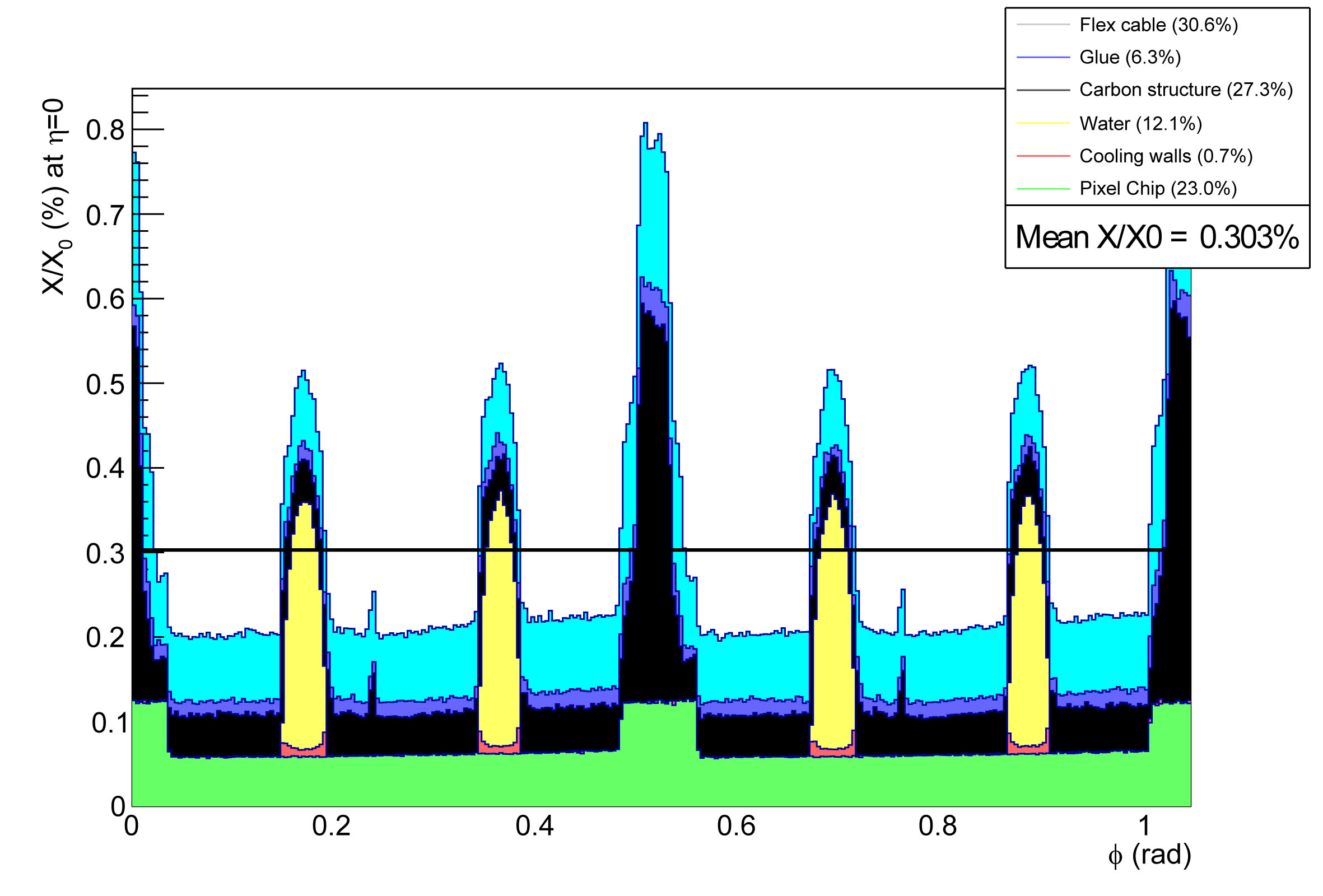}}
\caption{\label{mat22} The material budget distribution of Model 2.2.}
\end{figure}

\begin{figure}[th]
\centerline{\includegraphics[width=3.8in]{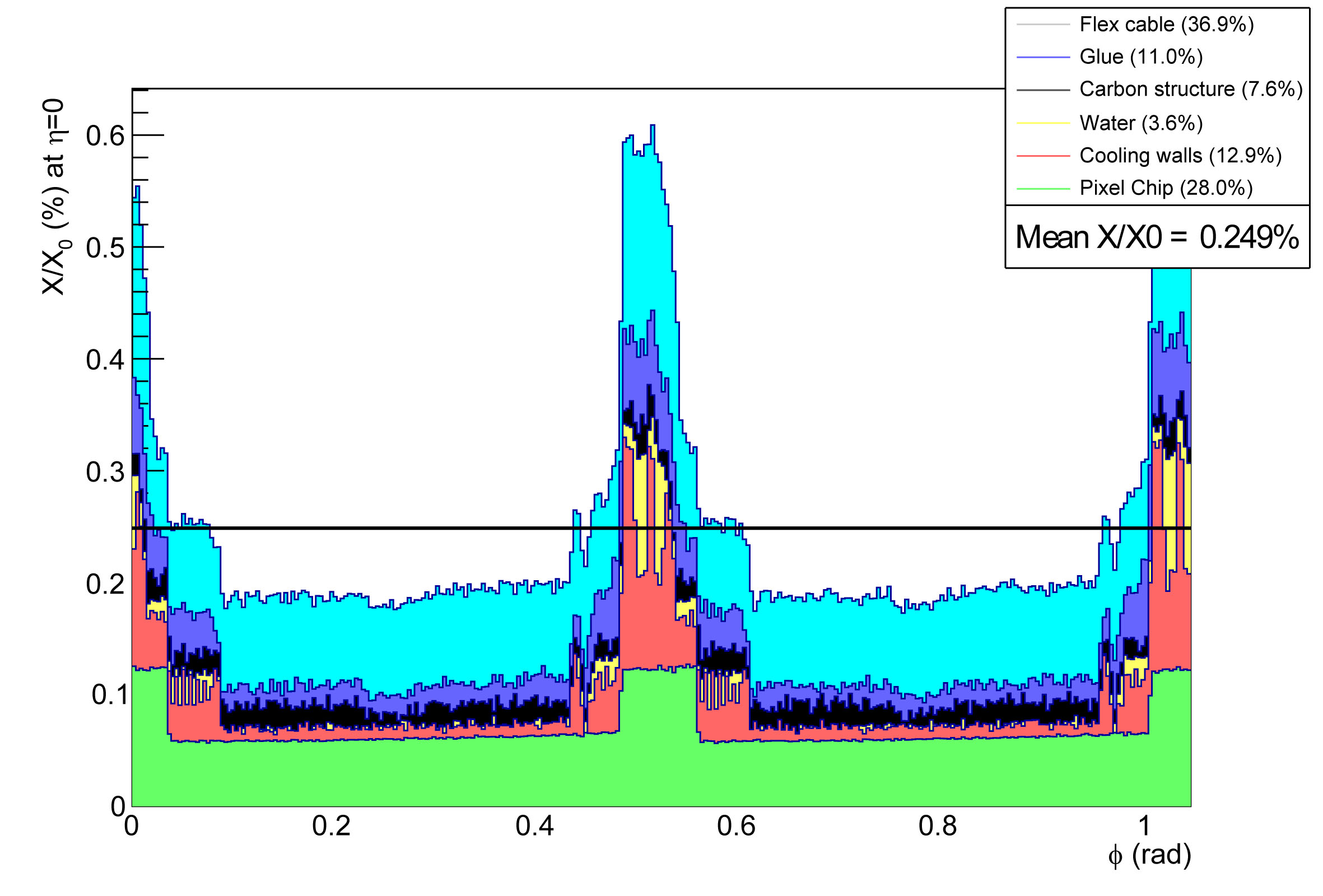}}
\caption{\label{mat3} The material budget distribution of Model 3.}
\end{figure}

The results of our calculations for material budget of different models of inner barrel are shown in Fig.~\ref{mat0}-\ref{mat3}. The highest peaks  represent the overlapping between each stave. In model 2.2, the location of the cooling pipes around the middle of the stave can help to enhance the thermal conductivity despite the higher material budget than model 0. The second highest peaks in Fig.~\ref{mat21} and Fig.~ \ref{mat22} are due to the polyimide cooling pipes filled with water. Instead of cooling pipes, using the microchannel filled water in model 1 or Freon in model 3 as a coolant can serve as a better average way to dissipate the heat from sensors~\cite{Rossegger201340}. However, model 1 has a higher material budget than model 2. While in model 3, the silicon micro channel is still in the R$\&$D phase and not ready to be implemented yet.

A summary of material budget calculation for all stave prototypes are given in Table~\ref{sum}.

\begin{table}[pt]
\tbl{\label{sum} Expected overall material budget obtained from ALIROOT simulation for different stave prototypes compared to the possible theoretical calculation in CDR~\cite{musa}.}
{\begin{tabular}{@{}ccc@{}} \toprule
Stave prototype & $X/X_0$ [CDR] & $X/X_0$ [ALIROOT]\\
 & ($\%$) & ($\%$)\\ \colrule
Model 0 & 0.26 & 0.284\\
Model 1 & 0.30 & 0.334\\
Model  &  & \\
\quad 2.1 & 0.31 & 0.344\\
\quad 2.2 & \quad - & 0.303\\
Model 3 & \quad - & 0.249\\ \botrule
\end{tabular}}
\end{table}

\section{Outer barrel stave}

\begin{figure}[htbp]
\centering 
\includegraphics[width=3.5in]{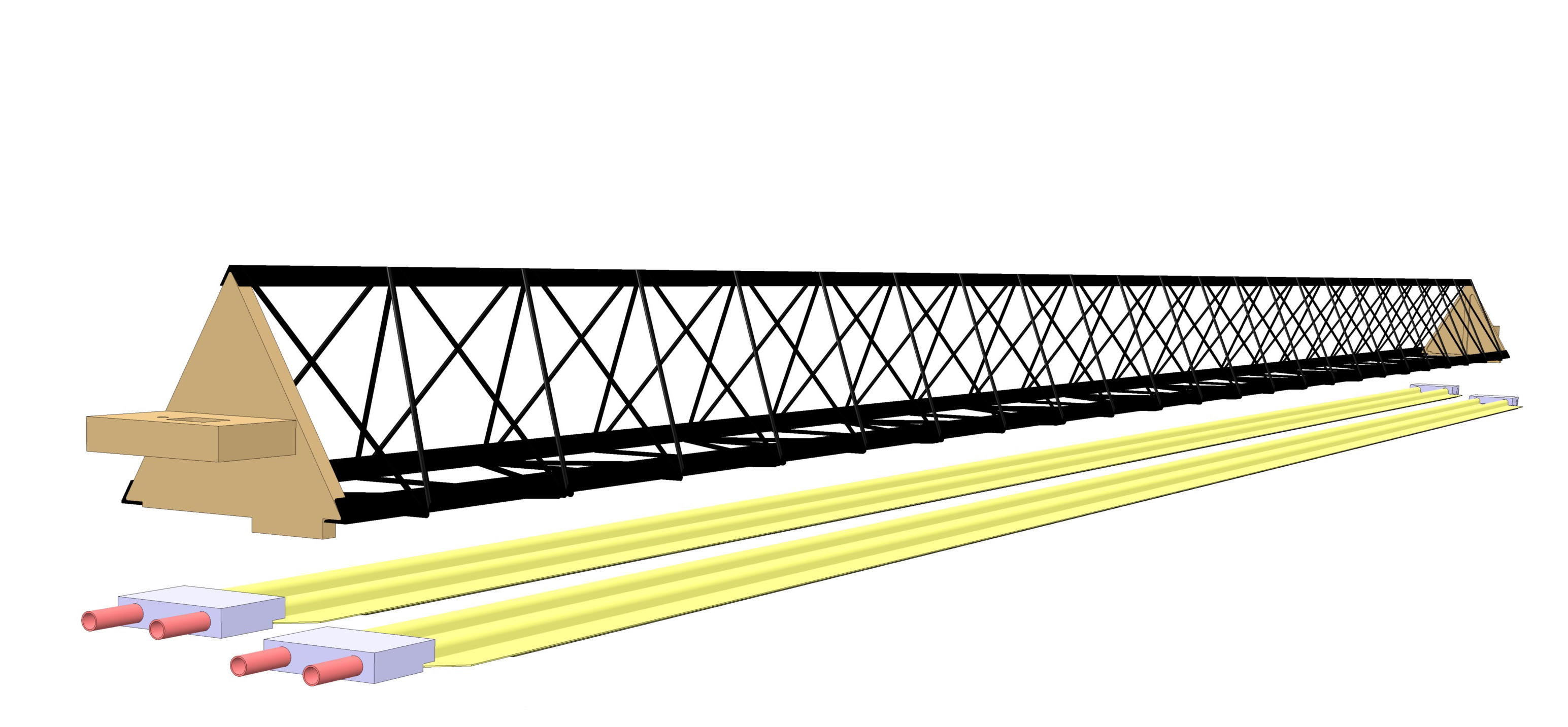}
\caption{\label{outer} Schematic layout of the mechanical and cooling structure of the outer barrel stave.~\cite{abelev}.}
\end{figure}

The conceptual design of the outer barrel stave is similar to the inner barrel. However, the bottom of the supporting part is split longitudinally into two half-staves on the azimuthal direction, i.e. along the Phi angle as shown in Fig.~\ref{outer}. In each half-staves consist of seven sensor chips with integrated cooling pipes and cold plates. These two pipes have 2.67 mm inner diameter and are filled with water.

The staves of the outer barrels have been designed to achieve the required stiffness and thermal properties as expected in inner barrels. Several components, similar to inner barrel, are used to prototype the outer barrel stave with seven-pixel chips. The estimate of the outer barrels stave to the material budget contribution are reported in Table~\ref{outtab}.

\begin{table}[pt]
\tbl{\label{outtab} List of the outer stave components and their thickness and the estimated contributions to the material budget~\cite{abelev}.}
{\begin{tabular}{@{}cccccc@{}} \toprule
Stave element & Component & Material & Thickness & $X/X_0$ & $X/X_0$\\
& &  & [$\upmu$m] & [cm] & [$\%$] \\ \colrule
\textbf{Module} & FPC Metal layers & Aluminium & 50 & 8.896 & 0.056\\ 
& FPC Insulating layers & Polyimide & 100 & 28.41 & 0.035\\
& FPC Insulating layers & Polyimide & 100 & 28.41 & 0.035\\
& Module plate & Carbon fibre & 120 & 26.08 & 0.046\\
& Pixel Chip & Silicon & 50 & 9.369 & 0.053\\
& Glue & Eccobond 45 & 100 & 44.37 & 0.023\\
\hline
\textbf{Power Bus} & Metal layers & Aluminium & 200 & 8.896 & 0.225\\ 
& Insulating layers & Polyimide & 200 & 28.41 & 0.070\\
& Glue & Eccobond 45 & 100 & 44.37 & 0.023\\
\hline
\textbf{Cold Plate} & & Carbon fleece & 40 & 106.80 & 0.004\\ 
& & Carbon paper & 30 & 26.56 & 0.011\\
& Cooling tube wall & Polyimide & 64 & 28.41 & 0.013\\
& Cooling fluid & water & & 35.76 & 0.105\\
& Carbon plate & Carbon fibre & 120 & 26.08 & 0.046\\
& Glue & Eccobond 45 & 100 & 44.37 & 0.023\\
\hline
\textbf{Space Frame} & & Carbon rowing & & & 0.080\\ 
\hline
\textbf{Total} & & & & & \textbf{0.813}\\ \botrule
\end{tabular}}
\end{table}

\begin{table}[pt]
\tbl{\label{element} Table of element composition of all the materials used in IB and OB stave.}
{\begin{tabular}{@{}cccc@{}} \toprule
Mixture material & Density g/cm$^3$ & Element & $\%$ Weight\\ \colrule
Kapton & 1.42 & H & 0.0226362\\ 
& & C & 0.69113\\
& & N & 0.07327\\
& & O & 0.209235\\
\hline
Air & 1.20479E-3 & C & 0.000124\\ 
& & N & 0.755267\\
& & O & 0.231781\\
& & Ar & 0.012827\\
\hline
Water & 1.0 & H & 0.111894\\ 
& & O & 0.888106\\
\hline
Flex cable & 1.0 & C & 0.520089\\ 
& & H & 0.019839\\
& & N & 0.055137\\
& & O & 0.157399\\
& & Al & 0.247536\\ \botrule
\end{tabular}}
\end{table}

\begin{table}[pt]
\tbl{\label{fiber} Materials and property used in IB and OB stave.}
{\begin{tabular}{@{}cc@{}} \toprule
SMaterial & Density g/cm$^3$\\ \colrule
Glue (C) & 1.93/2.015\\ 
K13D2U 2K (C) & 1.643\\ 
M60J 3k (C) & 2.21\\ 
FGS 003 (C) & 1.6\\ 
T300 (C) & 1.725\\ 
C-fleece (C) & 0.4\\ \botrule
\end{tabular}}
\end{table}

\begin{figure}[th]
\centerline{\includegraphics[width=5.in]{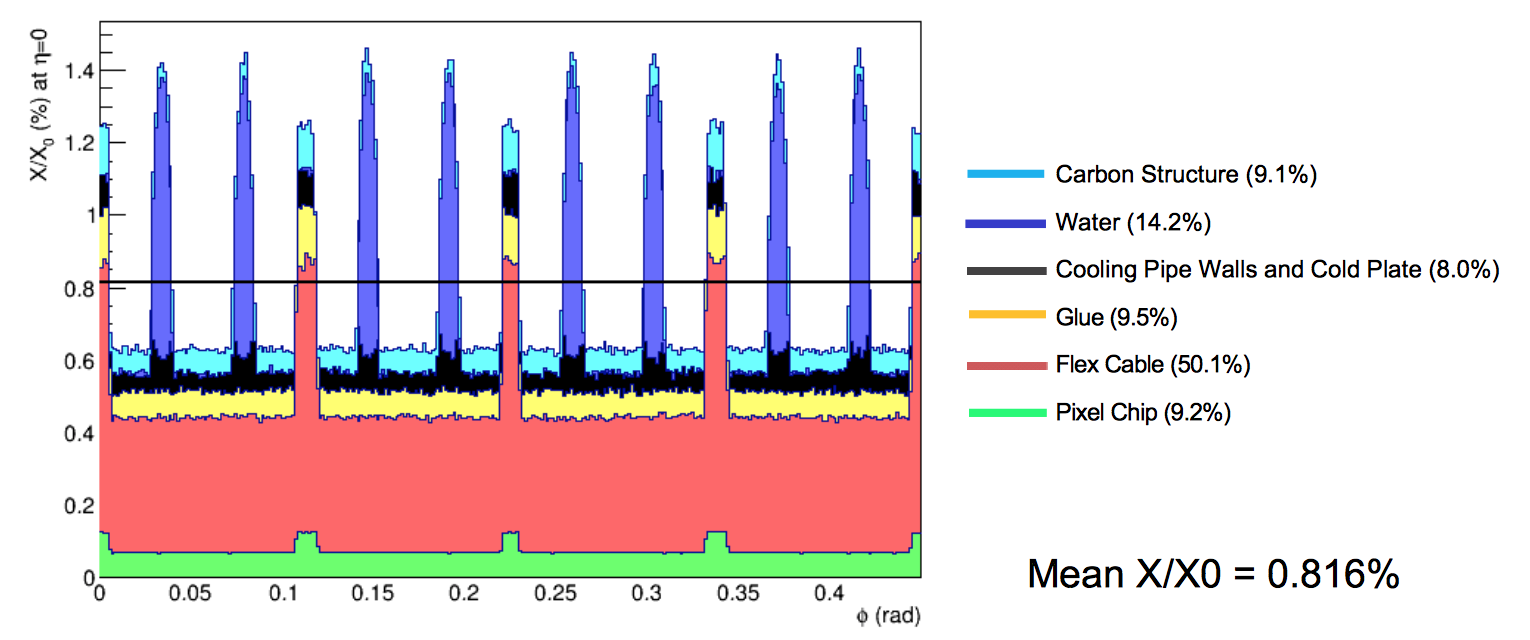}}
\caption{\label{matout} The material budget distribution of outer barrel prototype. The highest peaks correspond to the polyimide cooling pipes, filled of water, embedded in the cold plate~\cite{abelev}.}
\end{figure}

The results of the simulation of the material distribution across the outer barrel staves are shown in Fig.~\ref{matout}. The half-staves are partially superimposed surround the detector, thus giving rise to the peaks around 1.25$\% X_0$. The highest peaks are due to the polyimide cooling pipes, filled with water, embedded in the cold plate. The estimated overall material budget is within reach of the expected 0.8$\%$. 

\section{Conclusions}

The ITS upgrade design is aimed to improve the tracking efficiency of ALICE Detector and reduce the statistical uncertainty for low transverse momentum scattering. The construction of stave prototypes and the use of new cooling system will achieve a reduction of material budget for the new ITS. Presently, the ALIROOT simulation shows encouraging results on the material budget estimation. The results indicate the possibility of reducing the material budget to 0.3$\%$ per layer for Model 2.2 and 3. The simulation performed so far guarantee the conceptual design used in new ITS will qualify for ALICE Upgraded requirement. 

The final decision for the inner barrel stave structure goes for the revised model 2.2 or know as the current model 4. Currently, the stave production is ongoing to final assembly of the detector module.\cite{abelev}

\section*{Acknowledgements}

We would like to thank L. Musa and also the ALICE ITS upgrade team for their assistance during at CERN. This work was supported by Suranaree University of Technology (SUT), The office of the higher Education Commission under NRU project of Thailand (SUT-PhD/07/2556), National Research Council of Thailand (SUT1-105-58-24-03) and National Science and Technology Development Agency (CPMO P-15-50416). C. Kobdaj and Y. Yan acknowledge support from SUT and CHE-RNU project (SUT-COE: High Energy Physics $\&$ Astrophysics).

\end{document}